\documentclass[oneside,a4paper,reqno]{amsart}
\usepackage{amsmath}
\usepackage{amssymb}
\usepackage[mathscr]{eucal}
\usepackage{graphics}
\newcommand{\mat}[3]{	\left \langle  \vphantom{#2 #3}   #1   
                        \right|    					\, #2\,   
                        \left|    \vphantom{#2 #1} #3   
                        \right \rangle
                        								}
\newcommand{\bmat}[3]{\bigl \langle   #1 \bigr| \, #2\, \bigl|     #3  
\bigr \rangle}

\newcommand{\ket}[1]{		\left| #1 \right>  }
\newcommand{\bket}[1]{		\bigl| #1 \bigr> }
\newcommand{\bra}[1]{		\left< #1 \right|  }
\newcommand{\bbra}[1]{		\bigl< #1 \bigr|}

\newcommand{\overlap}[2]{	\left \langle    #1 \vphantom{#2 } \,
                        \right| \left.   #2 \vphantom{#1}
                        \right \rangle	}
\newcommand{\boverlap}[2]{ 
																								\bigl \langle #1 \, \bigr| \bigl. #2 \bigr \rangle
                        									}

\newcommand{\RErr}{     \Delta_{\mathrm{ei}}  }
\newcommand{\PErr}{     \Delta_{\mathrm{ef}}  }
\newcommand{\Dist}{     \Delta_{\mathrm{d}}  }
\newcommand{\am}{\hat{\mathbf{S}}}
\newcommand{\ams}{\hat{S}}
\newcommand{\amv}{\mathbf{S}}
\newcommand{\amvs}{S}

\newcommand{\ptv}{\mathbf{n}}
\newcommand{\pt}{\hat{\mathbf{n}}}
\newcommand{\pts}{\hat{n}}
\newcommand{\mpt}{\hat{\boldsymbol{\mu}}}
\newcommand{\mptv}{\boldsymbol{\mu}}
\newcommand{\mpts}{\hat{\mu}}
\newcommand{\mptvs}{\mu}
\newcommand{\av}{\hat{\xi}}
\newcommand{\avv}{\xi}
\newcommand{\apst}{\ket{\chi_{\rm ap}}}
\newcommand{\ap}{\chi_{\rm ap}}

\newcommand{\pdf}{\rho_{\rm val}}
\newcommand{\rf}{\eta_{\rm i}}
\newcommand{\pf}{\eta_{\rm f}}
\newcommand{\syf}{\eta_{\rm d}}
\newcommand{\ei}{\Delta_{\rm e i}}
\newcommand{\ef}{\Delta_{\rm e f}}
\newcommand{\dis}{\Delta_{\rm d}}
\newcommand{\Top}{\hat{T}(\ptv,\avv)}
\newcommand{\TopD}{\hat{T}^{\dagger}(\ptv,\avv)}
\newcommand{\Sop}{\hat{E}(\ptv)}
\newcommand{\Topm}{\hat{T}(\mptv,\avv)}
\newcommand{\TopmD}{\hat{T}^{\dagger}(\mptv,\avv)}
\newcommand{\Sopm}{\hat{E}(\mptv)}
\newcommand{\cov}{\mathrm{cv}}
\newcommand{\cnv}{\mathrm{cn}}
\newcommand{\proj}{\Pi_{2 s}(\ptv,\ptv')}
\DeclareMathOperator{\Tr}{Tr}
\begin{document}\begin{titlepage}
\begin{center}
\bfseries
OPTIMAL MEASUREMENTS OF SPIN DIRECTION
\end{center}
\vspace{1 cm}
\begin{center}
D M APPLEBY
\end{center}
\begin{center}
Department of Physics, Queen Mary and
		Westfield College,  Mile End Rd, London E1 4NS, UK
 \end{center}
\vspace{0.5 cm}
\begin{center}
  (E-mail:  D.M.Appleby@qmw.ac.uk)
\end{center}
\vspace{0.75 cm}
\vspace{1.25 cm}
\begin{center}
\textbf{Abstract}\\
\vspace{0.35 cm}
\parbox{10.5 cm }{ The  accuracy
of  a measurement of the spin direction of a spin-$s$ particle
is characterised, for arbitrary half-integral
$s$.  The disturbance caused by the
measurement is also characterised.  
The approach is based on that taken in several
previous papers concerning joint measurements of position
and momentum.  As in those papers, a distinction is made between
the errors of retrodiction and prediction.  Retrodictive and
predictive error relationships are derived.  The 
POVM describing the outcome of a maximally accurate measurement
process is investigated.  It is shown that, if the
measurement is retrodictively optimal, then the distribution of
measured values is given by the initial state
$\mathrm{SU}(2)$ $Q$-function.    If
the measurement is predictively optimal, then the distribution
of measured values is related to the final state
$\mathrm{SU}(2)$ $P$-function.  The general form of the unitary
evolution operator producing an optimal  measurement is
characterised.
                      }
\end{center}
\vspace{1 cm}
\begin{center}
Report no.  QMW-PH-99-18
\end{center}
\end{titlepage}
\title{Bohmian Velocity post-Decoherence}
\section{Introduction}
\label{sec:  intro}
In a recent series of
papers~\cite{self1,self2a,self2b,self3,self2c} we analysed the
concept of experimental accuracy, as it applies to simultaneous
measurements of position and
momentum~\cite{Arthurs,Peres1,Busch,Schroeck,Leonhardt}.  The
purpose of this paper is to give a similar analysis for
measurements of spin direction. 

There have been a number of previous discussions of joint,
imperfectly accurate measurements of two 
(non-commuting) components of spin~\cite{TwoComp}. 
Measurements of spin 
direction---the kind of measurement considered in this
paper---have been discussed  by Busch and
Schroeck~\cite{BuschSpin}, Grabowski~\cite{Grabowski},
Peres~\cite{Peres1}, and Busch
\emph{et al}~\cite{Busch}.  In the
following we extend the work of these authors by giving an
analysis of the measurement errors, and of the conditions for a
measurement process to be optimal.  In particular, we will show
that  a measurement is retrodictively optimal if and only if
the distribution of measured values is given by the generalized
$Q$-function which is defined in terms of $\mathrm{SU}(2)$
coherent states~\cite{SpinCoh,Lieb,Perel,SpinCohB}
(corresponding to an analogous property of joint measurements
of position and momentum derived by Ali and
Prugove\v{c}ki~\cite{Ali}, and proved under less
restrictive conditions in Appleby~\cite{self3}).

This result provides us with some further insight into the
physical significance of the $\mathrm{SU}(2)$ $Q$-function. 
It also has a bearing on the problem of state reconstruction. 
Amiet and Weigert~\cite{Amiet1,Amiet2} have recently shown how,
by making measurements of a single spin component for
sufficiently many  differently oriented Stern-Gerlach
apparatuses, one can calculate the corresponding values of the 
$\mathrm{SU}(2)$
$Q$-function, and thereby
reconstruct the density matrix.  The fact that a retrodictively
optimal measurement of spin direction has the $Q$-function as
its distribution of measured values suggests an alternative
approach to the problem of state reconstruction:  for it means
that one can reconstruct the density matrix from the
statistics of a single run of measurements,
performed on a single apparatus.  The fact that
measurements whose outcome is described the
$Q$-function have this property of informational completeness
has been stressed by Busch and Schroeck~\cite{BuschSpin} (also
see Busch~\cite{Complete}, Busch
\emph{et al}~\cite{Busch} and Schroeck~\cite{Schroeck}).

Retrodictively optimal joint measurements of
position  and momentum~~\cite{self3,Ali} give rise
to the ordinary Husimi or
$Q$-function~\cite{Husimi,Hillery,Lee}, and so they also have
the property of informational
completeness~\cite{Busch,Schroeck,Complete,Naka}, at least in
principle.  However, the practical usefulness of this fact is
somewhat restricted, due to the amplification of statistical
errors which occurs when one attempts to perform the
reconstruction starting from real experimental
data~\cite{Leonhardt,StatAmp}.  No such difficulty arises in
the case of measurements of spin direction, due to the fact
that the state space is finite dimensional.

We now outline the approach taken in the remainder of this
paper.   We consider a system consisting of a single spin, with
angular momentum operator $\am$ satisfying the usual commutation
relations
$\bigl[\ams_{a},\ams_{b}\bigr]=i
\sum_{c=1}^{3}\epsilon_{abc}\ams_{c}$ (with units chosen  such
that
$\hbar=1$).  We take it that
$\ams^2=s(s+1)$ for some arbitrary, but fixed  half-integer
$s$.

The components of
$\am$ are non-commuting, so they cannot all be simultaneously
measured with perfect precision.  However, they can all be
measured with a less than perfect degree of accuracy.  In order
to do so one can use the same kind of procedure which is
employed in the Arthurs-Kelly
process~\cite{self2a,Arthurs,Peres1,Busch,Schroeck,Leonhardt}: 
that is, one can  couple the non-commuting observables of
interest---the components of 
$\am$---to another set of ``pointer'' or ``meter'' observables
which do commute, and whose values may therefore be 
simultaneously determined with arbitrary precision.

The question we then have to decide is how to choose the
pointer observables.  The observables to be
measured satisfy the constraint $\ams^2=s(s+1)$, where $s$ is
fixed.  Consequently, one might take the view that the
magnitude of the spin vector is already known, and that all
that needs to be measured is its direction.
This suggests that the pointer observables should be taken to
be the (commuting) components of a unit vector $\pt$, satisfying
the constraint $\pts^2=1$.  The direction of $\pt$
measures the direction of $\am$.  We will refer to this as a
type 1 measurement.  Such measurements are discussed in 
Sections~\ref{sec:  POVM}--\ref{sec:  CompOpt}.

There is another possibility:  for one could take the pointer
observables to be the three \emph{independent}, commuting 
components of a vector
$\mpt$, no constraint being placed on the squared modulus
$\mpts^2$.  The value of $\ams_1$ (respectively $\ams_2$, 
$\ams_3$) is measured by $\mpts_1$ (respectively $\mpts_2$,
$\mpts_3$).  We will refer to this as a type 2 measurement. 
Such measurements are discussed in Section~\ref{sec:  type2}.

We begin our analysis in Section~\ref{sec:  POVM}, by
characterising the POVM (positive operator valued measure)
describing the outcome  of an arbitrary
type 1 measurement process.  

In 
Section~\ref{sec:  AccDis} we  characterise the
accuracy of and disturbance caused by a type 1 measurement
process.  Our definitions are based on those given  in
Appleby~\cite{self1,self2b}, for simultaneous measurements of
position and momentum.  In particular, we are led to make a
distinction between two different kinds of accuracy, which we
refer to as retrodictive and predictive.

After giving, in Section~\ref{sec:  CohSte}, a brief summary of
the relevant features of the theory of $\mathrm{SU}(2)$ coherent
states we go on, in Section~\ref{sec:  RetOpt}, to describe
retrodictively optimal type 1 measurements.  We establish a
bound on the retrodictive accuracy.  We  define a
retrodictively optimal measurement to be a measurement which
(1) achieves the maximum possible degree of retrodictive
accuracy, and which  (2) is isotropic (in a sense to be
explained).  We then show that the necessary and sufficient
condition for the measurement to be retrodictively optimal is
that the distribution of measured values be given by the
initial state $\mathrm{SU}(2)$ $Q$-function.

In Section~\ref{sec:  PreOpt} we establish a bound on the
predictive accuracy of a type 1 measurement.  We derive a
necessary and sufficient condition for this bound to be
achieved, in which case we say that the measurement is
predictively optimal.  We show that the distribution of
measured values is then related to the final  state 
$\mathrm{SU}(2)$ $P$-function.

In Section~\ref{sec:  CompOpt} we consider completely
optimal type 1 measurement processes---\emph{i.e.} processes
that are both retrodictively  and predictively optimal.   We
give the general form of the unitary evolution operator
describing such a process.

Finally, in Section~\ref{sec:  type2}, we consider type 2
measurements.  We define the retrodictive and predictive
errors of such measurements, and establish bounds which
the errors  must satisfy.  We then show that, in the limit
as a type 2 measurement tends to optimality (retrodictive or
predictive), it more and more nearly approaches an optimal type
1 measurement (with the replacement
$s^{-1}\mpt \rightarrow
\pt$).  It follows that, in so far as the aim is to maximise
the measurement accuracy,  type 2 measurements have no
advantages.
\section{Type 1 Measurements:  POVM}
\label{sec:  POVM}
The purpose of this section is to characterise the POVM
(positive operator valued
measure)~\cite{Peres1,Busch,Schroeck,Kraus,Peres2}
describing the outcome of an arbitrary type 1 measurement.

We take a type 1 measurement to consist of a process in which
the system, with $2s+1$ dimensional state space
$\mathscr{H}_{\rm sy}$, is coupled to a measuring apparatus,
with state space $\mathscr{H}_{\rm ap}$.  The interaction
commences at a time $t=t_{\rm i}$ when system$+$apparatus are
in the product state $\ket{\psi\otimes\ap}$, where
$\ket{\psi}\in
\mathscr{H}_{\rm sy}$ is the initial state of
the system and $\apst\in
\mathscr{H}_{\rm ap}$ is the initial state of the apparatus. It
ends after a finite time interval at
$t=t_{\rm f}$ when system$+$apparatus are in the state
$\hat{U}\ket{\psi\otimes\ap}$, where $\hat{U}$ is the unitary
evolution operator describing the measurement interaction.  

It should be stressed that this description is quite general. 
In particular, we are not making an impulsive approximation. 
Nor are we assuming that the interaction Hamiltonian is large in
comparison with  the Hamiltonians describing the system and
apparatus separately.    The only
substantive assumption is the statement that 
system$+$apparatus are initially in a product state (so that
they are initially uncorrelated).

It should be noted that $\ket{\psi}$ is arbitrary, since the
system might initially be in any state $\in \mathscr{H}_{\rm
sy}$. 
On the other hand $\apst$ is fixed, since
we assume that initially the apparatus is always in the same
``zeroed'' or ``ready'' state.

As explained in Section~\ref{sec:  intro}, we 
take it that the result of the measurement is specified by the
recorded values of three commuting pointer observables
$\pt=(\pts_1,\pts_2,\pts_3)$, satisfying the constraint
$\sum_{r=1}^{3}\pts_r^2=1$ (so that there are only two pointer
degrees of freedom).  However, a measuring instrument does not
usually consist of some pointers, and nothing else.  We
therefore allow for the existence of $N$ additional apparatus
observables $\av=(\av_1,\dots,\av_N)$ which, together with the
components of $\pt$, constitute a complete commuting set.  The 
eigenkets $\ket{\ptv,\avv}$ thus provide an orthonormal basis
for
$\mathscr{H}_{\rm ap}$.

The operator $\hat{U}$ specifies the final state of
system$+$apparatus given \emph{any} initial state
$\in \mathscr{H}_{\rm sy}\otimes\mathscr{H}_{\rm ap}$. 
However, we are only interested in initial states of the very
special form
$\ket{\psi\otimes \ap}$, where $\apst$ is fixed. In other
words, the operator  $\hat{U}$ provides us with much more
information than we actually need.  It turns out that all the
quantities which are relevant to the argument of this paper can
be expressed in terms of the
operator $\Top$, defined
by~\cite{Busch,Schroeck,Kraus,Peres2}
\begin{equation}
  \Top = \sum_{m_1,m_2=-s}^{s}
             \bigl(\bra{m_1}\otimes\bra{\ptv,\avv}\bigr)
               \hat{U}
          \bigl(\ket{m_2}\otimes\apst\bigr)
          \ket{m_1}\bra{m_2}
\label{eq:  TDef}
\end{equation}
where $\ket{m}$ denotes the eigenket of $\ams_3$ with
eigenvalue $m$ (in units such that $\hbar=1$).  The operator
$\Top$ is more convenient to work with because, unlike
$\hat{U}$, it only acts on the system state space
$\mathscr{H}_{\rm sy}$.

The significance of the operator $\Top$ is that it describes
the change in the state of the system which is caused by the
measurement process~\cite{Busch,Schroeck,Kraus,Peres2}
(\emph{i.e.}\ it describes the
operation~\cite{Busch,Schroeck, Kraus} induced by the
measurement).  In fact, suppose that the measurement is
non-selective (meaning that the final value of
$\ptv$ is not recorded, so that there is no ``collapse''), and
let
$\hat{\rho}_{\rm f}$ be the reduced density matrix describing
the final state of the system.  It is then readily verified that
\begin{equation}
  \hat{\rho}_{\rm f}
=  \int d\ptv \, d \avv \;
   \Top\,
       \ket{\psi}\bra{\psi} \,
   \TopD
\label{eq:  rhof}
\end{equation}
where $d \ptv$ denotes the usual measure on the
unit $2$-sphere:  in terms of spherical polars
$d \ptv = \sin \theta d\theta d \phi$.

Let $\pdf(\ptv)$ be the probability density function
describing the distribution of measured values:
\begin{equation}
  \pdf(\ptv)
=  \sum_{m =-l}^{l} \int d\avv \,
     \Bigl| \bigl(\bra{m}\otimes \bra{\ptv,\avv}\bigr)
            \hat{U}
               \bigl(\ket{\psi}\otimes \apst\bigr)
     \Bigr|^2
\label{eq:  pdfDef}
\end{equation}
$\pdf(\ptv)$ can also be expressed in terms of the operators
$\Top$.  In fact,
define~\cite{Busch,Schroeck,Kraus,Peres2}
\begin{equation}
  \Sop
=  \int d \avv \, \TopD \Top
\label{eq:  SDef}
\end{equation}
Then
\begin{equation}
  \pdf(\ptv) = \bmat{\psi}{\Sop}{\psi}
\label{eq:  val}
\end{equation}
We see from this that $\Sop d\ptv$ is the POVM describing the
measurement outcome.  In particular
\begin{equation*}
  \Sop \ge 0
\end{equation*}
for all $\ptv$ and
\begin{equation}
  \int d \ptv \, \Sop = 1
\label{eq:  SNorm}
\end{equation}

Until now we  have been assuming that the system is initially
in a pure state.  If the system is initially in the mixed state
with density matrix $\hat{\rho}_{\rm i}$ we have, in place of 
Eqs.~(\ref{eq:  rhof}) and~(\ref{eq:  val}),
\begin{equation}
 \hat{\rho}_{\rm f}
= \int d \ptv \, d\avv \,
  \Top \,  \hat{\rho}_{\rm i} \, \TopD
\label{eq:  rhofMix}
\end{equation}
and
\begin{equation}
  \pdf(\ptv)
=  \Tr \left(\Sop \, \hat{\rho}_{\rm i}\right)
\label{eq:  rhoValTermsRhoi}
\end{equation}
Eq.~(\ref{eq:  rhofMix}) gives the final state  reduced
density for the system in the case when the measurement is
non-selective, so that the pointer position is not recorded. 
Suppose, on the other hand, that the final pointer position is
recorded to be in the subset $\mathscr{R}$ of the unit
$2$-sphere.  Then $\hat{\rho}_{\rm f}$ is given by
\begin{equation}
  \hat{\rho}_{\rm f}
= \frac{1}{p_{\mathscr{R}}}
  \int_{\mathscr{R}} d\ptv \int d\avv \,
  \Top \, \hat{\rho}_{\rm i} \, \TopD
\label{eq:  rhofTermsT}
\end{equation}
where $p_{\mathscr{R}}$ is the probability of finding
$\ptv \in \mathscr{R}$:
\begin{equation*}
  p_{\mathscr{R}} = \int_{\mathscr{R}} d\ptv \, \pdf(\ptv)
\end{equation*}
\section{Type 1 Measurements:  Accuracy and Disturbance}
\label{sec:  AccDis}
The purpose of this paper is to establish the form of the
operators $\Top$ and $\Sop$ when the measurement is optimal. 
In order to give a precise definition of what ``optimal'' means
in this context, we first need to define a concept
of measurement accuracy;  which is the problem addressed in
this section.  We also discuss how to quantify the degree to
which the system is disturbed by the measurement process.

The approach we take is based on the approach taken in
Appleby~\cite{self1,self2b}, to the problem of defining the
accuracy of and disturbance caused by a simultaneous
measurement of position and momentum.
We thus work in terms of the Heisenberg picture.  

Let
$\am_{\rm i} = \am$ and $\pt_{\rm i} = \pt$ be the initial
values of the Heisenberg spin and pointer observables at the
time $t_{\rm i}$, when the measurement interaction begins; and
let $\am_{\rm f} = \hat{U}^{\dagger}\am\hat{U}$ and $\pt_{\rm
f} = \hat{U}^{\dagger}\pt\hat{U}$ be the final values of these
observables at the time $t_{\rm f}$, when the measurement
interaction ends. Let
$\mathscr{S}_{\rm sy}\subset\mathscr{H}_{\rm sy}$ be the unit
sphere in the system state space.  We then define the
retrodictive fidelity $\rf$ by
\begin{equation}
\rf = \inf_{\ket{\psi}\in\mathscr{S}_{\rm sy}}
\Bigl(\bmat{\psi\otimes\ap}{\tfrac{1}{2}\bigl(\pt_{\rm f}
\cdot \am_{\rm i}+
\am_{\rm i}
\cdot \pt_{\rm f}\bigr)}{\psi\otimes \ap}\Bigr)
\label{eq:  rfDef}
\end{equation}
and the 
predictive fidelity $\pf$ by
\begin{align}
\pf 
& = \inf_{\ket{\psi}\in\mathscr{S}_{\rm sy}}
\Bigl(\bmat{\psi\otimes\ap}{\tfrac{1}{2}\bigl(\pt_{\rm f} \cdot
\am_{\rm f}+
\am_{\rm f}
\cdot \pt_{\rm f}\bigr)}{\psi\otimes\ap}\Bigr)
\notag
\\ & =
\inf_{\ket{\psi}\in\mathscr{S}_{\rm sy}}
\Bigl(\bmat{\psi\otimes\ap}{\pt_{\rm f} \cdot \am_{\rm
f}}{\psi\otimes\ap}\Bigr)
\label{eq:  pfDef}
\end{align}
(where we have used the fact that the components of 
$\pt_{\rm f}$ and $\am_{\rm f}$ commute).  
It should be noted that the concept of fidelity employed here
is somewhat different from the concept of fidelity which is
employed in discussions of cloning and state estimation ($\rf$
and
$\pf$ are defined in terms of  scalar products of
observables, rather than scalar products of  states).

We also define the quantity $\syf$ by
\begin{equation}
\syf = \inf_{\ket{\psi}\in\mathscr{S}_{\rm sy}}
\Bigl(\bmat{\psi\otimes\ap}{\tfrac{1}{2}\bigl(\am_{\rm f} \cdot
\am_{\rm i}+
\am_{\rm i}
\cdot \am_{\rm f}\bigr)}{\psi\otimes\ap}\Bigr)
\label{eq:  syfDef}
\end{equation}

The intuitive basis for these definitions is most easily
appreciated if one thinks, temporarily, in classical terms.  If
interpreted classically
$\rf$ would represent the minimum expected degree of alignment
between the final pointer direction and the initial direction
of the spin vector.  In other words, it would quantify the
retrodictive accuracy of the measurement.  On the other hand,
$\pf$ would represent the minimum expected degree of alignment
between the final pointer direction and the final direction of
the spin vector:  it would therefore provide a quantitative
indication of the predictive accuracy.  Lastly,  $\syf$
would quantify the extent to which the measurement disturbs the
system, by changing the direction of the spin vector. 

Of course, $\pt_{\rm f}$, $\am_{\rm i}$, $\am_{\rm f}$ are in
fact quantum mechanical observables, and so the physical
interpretation of $\rf$, $\pf$ and $\syf$ needs to be justified
much more carefully.  Rather than proceeding directly, it will
be convenient first to relate these quantities to an
alternative characterisation of the measurement accuracy and
disturbance.  This will allow us to appeal to the arguments
given in Appleby~\cite{self1,self2b}, to justify our earlier
characterisation of the accuracy of and disturbance caused by a
simultaneous measurement of position and momentum.  It will
also be helpful in Section~\ref{sec:  type2},
when we compare type 1 and type 2 measurements.

In a type 1 measurement, the result of the measurement is a
direction, represented by the unit vector $\ptv$.  However, one
could extract from this information estimates of the
initial and final values of the spin vector itself by
multiplying
$\ptv$ by  suitable constants:  say 
$\zeta_{\rm i} \ptv$ as an estimate for
$\amv_{\rm i}$, and 
$\zeta_{\rm f} \ptv$ as an estimate for $\amv_{\rm f}$.  The
question then arises:  what are the best choices for these
constants?

To answer this question, consider the quantities
\begin{align*}
  \sup_{\ket{\psi }\in \mathrm{S}_{\rm sy}}
  \left( \bmat{\psi\otimes\ap}{
               \bigl|\zeta_{\rm i} \pt_{\rm f}-\am_{\rm i}
\bigr|^2}{\psi\otimes\ap}
  \right)
& = \zeta_{\rm i}^{2} - 2 \zeta_{\rm i} \rf + s(s+1)
\\
\intertext{and}
  \sup_{\ket{\psi }\in \mathrm{S}_{\rm sy}}
  \left( \bmat{\psi\otimes\ap}{
               \bigl|\zeta_{\rm f} \pt_{\rm f}-\am_{\rm f}
\bigr|^2}{\psi\otimes\ap}
  \right)
& = \zeta_{\rm f}^{2} - 2 \zeta_{\rm f} \pf + s(s+1)
\end{align*}
These expressions are minimised if we choose 
$\zeta_{\rm i} = \rf$,
$\zeta_{\rm f} = \pf$.  We accordingly define the maximal
rms error of retrodiction 
\begin{equation}
\ei \amvs
 = \left(\sup_{\ket{\psi }\in \mathrm{S}_{\rm sy}}
  \left( \bmat{\psi\otimes\ap}{
               \bigl|\rf \pt_{\rm f}-\am_{\rm i}
\bigr|^2}{\psi\otimes\ap}
  \right)
        \right)^{\frac{1}{2}}
= \left( s + s^2-\rf^2
  \right)^{\frac{1}{2}}
\label{eq:  RetErrA}
\end{equation}
and the maximal rms error of prediction
\begin{equation}
\ef \amvs
 = \left(\sup_{\ket{\psi }\in \mathrm{S}_{\rm sy}}
  \left( \bmat{\psi\otimes\ap}{
               \bigl|\pf \pt_{\rm f}-\am_{\rm f}
\bigr|^2}{\psi\otimes\ap}
  \right)
        \right)^{\frac{1}{2}}
= \left( s + s^2-\pf^2
  \right)^{\frac{1}{2}}
\label{eq:  PreErrA}
\end{equation}
We also define the maximal rms disturbance by
\begin{equation}
\dis \amvs
 = \left(\sup_{\ket{\psi }\in \mathrm{S}_{\rm sy}}
  \left( \bmat{\psi\otimes\ap}{
               \bigl| \am_{\rm f}-\am_{\rm i}
\bigr|^2}{\psi\otimes\ap}
  \right)
        \right)^{\frac{1}{2}}
= \sqrt{2}\left( s + s^2-\syf
  \right)^{\frac{1}{2}}
\label{eq:  DistA}
\end{equation}
Comparing these expressions with those given in
refs.~\cite{self1,self2b} it can be seen that 
$\ei \amvs$ plays the same role in relation to the kind of
measurement here considered as do the quantities 
$\RErr x$, $\RErr p$ in relation to  joint
measurements of position and momentum; that 
$\ef \amvs$ is the analogue of $\PErr x$, $\PErr p$; and that 
$\dis \amvs$ is the analogue of $\Dist x$, $\Dist p$.
A suitably modified version of
the argument given in Section 5 of ref.~\cite{self1} may then be
used to show that $\ei \amvs$ (and therefore $\rf$) describes
the retrodictive accuracy of the measurement; that $\ef \amvs$
(and therefore
$\pf$) describes the predictive accuracy; and that $\dis \amvs$
(and therefore $\syf$) describes the degree of disturbance
caused by the measurement.

Finally, we note that the quantities $\rf$, $\pf$ and $\syf$ can
be expressed in terms of the operators $\hat{T}(\ptv, \avv)$ and
$\hat{S}(\ptv)$ defined earlier.  In fact, 
comparing  Eqs.~(\ref{eq:  TDef})
and~(\ref{eq:  SDef}) with 
Eqs.~(\ref{eq:  rfDef}--\ref{eq:  syfDef}) one finds
{\allowdisplaybreaks
\begin{align}
 \rf
& = \inf_{\ket{\psi}\in\mathscr{S}_{\rm sy}}
  \left( \int d \ptv \, 
     \bmat{\psi}{\tfrac{1}{2}
       \bigl( \Sop \ptv \cdot \am + \ptv \cdot \am \, \Sop
       \bigr)}{\psi}
  \right)
\label{eq:  etaiTermsS}
\\
\pf
& = \inf_{\ket{\psi}\in\mathscr{S}_{\rm sy}}
  \left( \int d \ptv d\avv \, 
     \bmat{\psi}{ \TopD\, \ptv \cdot \am  \, \Top}{\psi}
  \right)
\label{eq:  etafTermsT}
\end{align}
and
\begin{multline}
\syf
 =\inf_{\ket{\psi}\in\mathscr{S}_{\rm sy}}
  \biggl( \int d \ptv d \avv \, 
  \sum_{a=1}^{3}
     \bbra{\psi} \tfrac{1}{2}
       \bigl( \TopD \, \ams_{a}\, \Top \, \ams_{a}
   \bigr.
   \biggr.
\\
   \biggl.
   \bigl.
   +
   \ams_{a}\, \TopD \,  \ams_{a} \, \Top 
        \bigr)\bket{\psi}
  \biggr)
\label{eq:  etadTermsT}
\end{multline}
}
\section{$\mathrm{SU}(2)$ Coherent States}
\label{sec:  CohSte}
The task we now face is to establish upper bounds on the
fidelities $\rf$,
$\pf$ (or, equivalently, lower bounds on the errors
$\ei \amvs$, $\ef \amvs$), and then to
establish the form of the operators
$\Top$,
$\Sop$ for which these bounds are achieved.  The theory of
$\mathrm{SU}(2)$ coherent states will play an important role
in the argument.   In order to fix notation we begin by
summarising the relevant parts of this theory.  For proofs of
the statements made in this section see
refs.~\cite{SpinCoh,Lieb,Perel,SpinCohB}.

For each unit vector $\ptv \in \mathbb{R}^3$ choose a vector
$\boldsymbol{\theta}_{\ptv}\in \mathbb{R}^3$ with the property
\begin{equation*}
  \exp\bigl[-i \boldsymbol{\theta}_{\ptv} \cdot \am \bigr] \,
  \ams_{3} \,
  \exp\bigl[i \boldsymbol{\theta}_{\ptv} \cdot \am \bigr]
= \ptv \cdot \am
\end{equation*}
Define
\begin{equation}
  \ket{\ptv,m}
=  \exp\bigl[ - i \boldsymbol{\theta}_{\ptv} \cdot
     \am \bigr]\ket{m}
\label{eq:  mnKetDef}
\end{equation}
where $\ket{m}$ is the normalized eigenvector of 
$\ams_{3}$ with eigenvalue $m$.
We then have
\begin{equation}
  \ptv \cdot \am \ket{\ptv,m} = m \ket{\ptv,m}
\label{eq:  CohSteEval}
\end{equation}
and
\begin{equation*}
  \frac{2s+1}{4 \pi}
  \int d\ptv \,
    \ket{\ptv,m}\bra{\ptv,m}
= 1
\end{equation*}
for all $m$.

We are especially interested in the states $\ket{\ptv,s}$. 
These are the minimum uncertainty states, for which
$\sum_{a=1}^{3} \bigl(\Delta \ams_{a}\bigr)^2=s$.  To denote
them we employ the abbreviated notation
\begin{equation}
  \ket{\ptv}=\ket{\ptv,s}
\label{eq:  nKetDef}
\end{equation}
The states $\ket{\ptv}$ so defined
$\in
\mathscr{H}_{\rm sy}$ and are  eigenvectors of  $\ptv \cdot
\am$.  They need to be carefully distinguished from the states
$\ket{\ptv, \avv}$ which $\in \mathscr{H}_{\rm ap}$ and 
are eigenvectors of  $\pt$.

Let $\hat{A}$ be any operator acting on $\mathscr{H}_{\rm sy}$. 
The covariant symbol corresponding to $\hat{A}$ is defined by
\begin{equation*}
  A_{\cov}(\ptv) = \bmat{\ptv}{\hat{A}}{\ptv}
\end{equation*}
The contravariant  symbol corresponding to $\hat{A}$ is
defined to be the unique function $A_{\cnv}$ for which
\begin{equation*}
  \hat{A} = \frac{2s+1}{4 \pi}
    \int d \ptv \,
        A_{\cnv}(\ptv) \ket{\ptv}\bra{\ptv}
\end{equation*}
and which satisfies
\begin{equation*}
    \int d\ptv' \proj A_{\cnv}(\ptv') = A_{\cnv}(\ptv)
\end{equation*}
where $\proj$ is the projection kernel
\begin{equation}
 \proj
= \sum_{j=0}^{2 s}\sum_{m=-j}^{j}
    Y^{\vphantom{*}}_{jm}(\ptv) Y^{*}_{jm}(\ptv')
= \sum_{j=0}^{2 s} \frac{2 j+1}{4 \pi} P_{j}(\ptv \cdot \ptv')
\label{eq:  Pi0Proj}
\end{equation}
In these expressions the $Y_{jm}$ are spherical harmonics and
the 
$P_{j}$ are Legendre polynomials.

It can be shown that, given any square integrable function $f$,
\begin{equation}
  \hat{A} 
= \frac{2s+1}{4 \pi}
    \int d\ptv \,f(\ptv) \ket{\ptv}\bra{\ptv}
\label{eq:  fCondA}
\end{equation}
if and only if
\begin{equation}
  \int d\ptv' \proj f(\ptv') = A_{\cnv} (\ptv)
\label{eq:  fCondB}
\end{equation}
for almost all $\ptv$.

The covariant (respectively contravariant) symbol of an
operator is often referred to as the $Q$ (respectively $P$)
symbol of that operator.  However, we will find it more
convenient to reserve this notation for the  symbols
corresponding specifically to the density matrix, scaled by a
factor
$(2s+1)/(4 \pi)$:
\begin{align}
  Q(\ptv) & = \frac{2s+1}{4 \pi}\rho_{\cov} (\ptv) \\
  P(\ptv) & = \frac{2s+1}{4 \pi}\rho_{\cnv} (\ptv) 
\label{eq:  PfncDef}
\end{align}
With this rescaling the $Q$ and $P$-functions satisfy the
normalisation condition
\begin{equation*}
  \int d\ptv \, Q(\ptv) = \int d\ptv \, P(\ptv) =1
\end{equation*}
In particular, $Q(\ptv)$ is a probability density function.  As
we will see, it is in fact the probability density function
describing the outcome of a retrodictively optimal type 1
measurement.

\section{Retrodictively Optimal Type 1 Measurements}
\label{sec:  RetOpt}
The purpose of this section is to investigate those processes
which maximise the retrodictive fidelity.  We begin by
establishing the following bound on $\rf$:
\begin{equation}
  \rf \le s
\label{eq:  etaiCondA}
\end{equation}
which, in view of Eq.~(\ref{eq:  RetErrA}), implies
\begin{equation}
  \ei \amvs \ge \sqrt{s}
\label{eq:  RetErrRelA}
\end{equation}
We will refer to Inequality~(\ref{eq:  RetErrRelA}) as the
retrodictive error relation.  It can be seen that it has the
same form as the ordinary uncertainty relation,
$\Delta \amvs \ge \sqrt{s}$.  It is the analogue, for the kind
of measurement here considered, of the inequality 
$\RErr x \, \RErr p \ge 1/2$ proved in ref.~\cite{self2b} for
joint measurements of position and momentum (in units such
that $\hbar=1$).

In order to prove this result we note that it follows from
Eqs.~(\ref{eq:  SDef}) and~(\ref{eq:  etaiTermsS}) that
\begin{equation*}
   (2s+1) \rf
\le  \int d \ptv d \avv \, 
     \Tr \bigl(  \ptv \cdot \am \, \TopD  \Top \bigr)
\end{equation*}
In view of Eqs.~(\ref{eq:  SDef})
and~(\ref{eq:  SNorm}) we also have
\begin{equation*}
  \int d\ptv d\avv \,
    \Tr \bigl( \TopD  \Top \bigr)
=  (2s +1)
\end{equation*}
Consequently
\begin{equation}
  \int d\ptv d\avv \,
     \Tr\bigl( (\rf-\ptv \cdot \am)
     \TopD \Top \bigr)
\le 0
\label{eq:  RetFidCondC}
\end{equation}
For each fixed $\ptv$ the kets $\ket{\ptv,m}$ defined by
Eq.~(\ref{eq:  mnKetDef}) constitute an orthonormal basis. 
We may therefore write
\begin{equation}
  \Top
=  \sum_{m,m'=-s}^{s}
      T_{m m'}(\ptv,\avv) \ket{\ptv,m}\bra{\ptv,m'}
\label{eq:  TExpand}
\end{equation}
for suitable coefficients $T_{m m'}$.  Substituting this
expression in Inequality~(\ref{eq:  RetFidCondC}) gives
\begin{equation}
  \sum_{m, m'=-s}^{s}
  \left(
  (\rf-m')
  \int d\ptv d\avv\,
   |T_{m m'}(\ptv,\avv)|^2
   \right)
\le 0
\label{eq:  RetFidCondD}
\end{equation}
Inequality~(\ref{eq:  etaiCondA}) is now immediate.

We next show that the retrodictive fidelity achieves its
maximum value $\rf = s$ if and only if $\Sop$ is of the form
\begin{equation}
  \Sop
=  \frac{2s+1}{4 \pi}g(\ptv) \ket{\ptv}\bra{\ptv}
\label{eq:  SforRfMax}
\end{equation}
for almost all $\ptv$, 
where $\ket{\ptv}$ is the state defined by Eq.~(\ref{eq: 
nKetDef}), and where $g$ is any function satisfying
\begin{equation}
  \int d\ptv' \, \proj g(\ptv') =1
\label{eq:  gCond}
\end{equation}[$\proj$ being the
projection kernel defined by Eq.~(\ref{eq:  Pi0Proj})].

In fact, setting $\rf=s$ in 
Inequality~(\ref{eq:  RetFidCondD}) gives
\begin{equation}
  \sum_{m, m'=-s}^{s}
  \left(
  (s-m')
  \int d\ptv d\avv\,
   |T_{m m'}(\ptv,\avv)|^2
   \right)
\le 0
\end{equation}
from which it follows that the coefficients $T_{m m'}$ must be
of the form 
\begin{equation*}
  T_{m m'}(\ptv,\avv)
= \left( \frac{2s+1}{4 \pi}\right)^{\frac{1}{2}}
\delta_{m' l} \, g_{m}(\ptv, \avv)
\end{equation*}
for almost all $\ptv$, $\avv$.  Substituting this expression
into Eq.~(\ref{eq:  TExpand}) gives
\begin{equation}
  \Top
= \left( \frac{2s+1}{4 \pi}\right)^{\frac{1}{2}}
  \ket{g(\ptv,\avv)}\bra{\ptv}
\label{eq:  RetOptTCondC}
\end{equation}
for almost all $\ptv$, $\avv$, where
\begin{equation*}
  \ket{g(\ptv,\avv)} = \sum_{m=-s}^{s} g_{m}(\ptv,\avv)
  \ket{\ptv,m}
\end{equation*}
Setting 
\begin{equation*}
 g(\ptv) = \int d\avv \, \bigl\|\ket{g(\ptv,\avv)} \bigr\|^2
\end{equation*}
and using Eq.~(\ref{eq:  SDef}), we deduce that $\Sop$ is of the form
specified by Eq.~(\ref{eq:  SforRfMax}).
It follows from Eqs.~(\ref{eq:  SNorm}),
(\ref{eq:  fCondA}) 
and~(\ref{eq:  fCondB}), and the fact that $\mathrm{id}_{\rm
\cnv}(\ptv)=1$, that the function
$g$ must satisfy
 Eq.~(\ref{eq:  gCond}).  This proves
that the condition represented by Eqs.~(\ref{eq:  SforRfMax})
and~(\ref{eq:  gCond}) is necessary.

Suppose, on the other hand, that $\Sop$ is given by
Eq.~(\ref{eq:  SforRfMax}), with $g$ satisfying Eq.~(\ref{eq: 
gCond}).  Using Eqs.~(\ref{eq:  etaiTermsS}), 
(\ref{eq:  fCondA}) 
and~(\ref{eq:  fCondB}) we deduce
\begin{equation*}
  \rf  =
  \inf_{\ket{\psi}\in\mathscr{S}_{\rm sy}}
  \left(\frac{2s+1}{4 \pi}\int d\ptv \,
     s g(\ptv) \bigl|\overlap{\ptv}{\psi}
     \bigr|^2
  \right)
 = s
\end{equation*}
which shows that the condition is also sufficient.

The condition $\rf =s$ is not, by itself, enough to determine
the distribution of measured values.  However, the requirement
that the retrodictive fidelity be maximised is not the only
property which it is natural to require of a measurement that
is to count as optimal.  It is also natural to require that the
measurement does not pick out any distinguished spatial
directions. We accordingly define an isotropic measurement to
be one which has the property that, if the initial system state
density matrix takes the rotationally invariant form
\begin{equation*}
  \hat{\rho}_{\rm i} = \frac{1}{2 s+1}
\end{equation*}
then the distribution of measured values is also rotationally
invariant:
\begin{equation*}
  \pdf(\ptv) = \frac{1}{4 \pi}
\end{equation*}
for all $\ptv$.  

We define a retrodictively optimal type 1 measurement process
to be an isotropic process for which the retrodictive fidelity
is maximal, $\rf=s$.  It is then straightforward to verify that
a type 1 measurement process is retrodictively optimal if and
only if $\Sop = (2s+1)/(4 \pi) \ket{\ptv}\bra{\ptv}$.  This is
the POVM which has previously been discussed by Busch and
Schroeck~\cite{BuschSpin}, and
others~\cite{Peres1,Busch,Schroeck,Grabowski}.

We see from Eq.~(\ref{eq:  rhoValTermsRhoi}) that the
measurement is retrodictively optimal if and only if the
distribution of measured values is given by
\begin{equation*}
  \pdf(\ptv) = Q_{\rm i} (\ptv)
\end{equation*}
for all $\ptv$, where $Q_{\rm i}$ is the $Q$-function
corresponding to the initial system state density matrix:
\begin{equation*}
  Q_{\rm i} (\ptv) 
= \frac{2s+1}{4 \pi}\mat{\ptv}{\hat{\rho}_{\rm i}}{\ptv}
\end{equation*}

In terms of the operator $\Top$, the necessary and
sufficient condition for a type 1 measurement to be
retrodictively optimal is that 
[see  Eq.~(\ref{eq:  RetOptTCondC})]
\begin{equation}
  \Top
= \left( \frac{2s+1}{4 \pi}\right)^{\frac{1}{2}}
  \ket{g(\ptv,\avv)}\bra{\ptv}
\label{eq:  RetOptTCondA}
\end{equation}
where $\ket{g(\ptv,\avv)}$ is any
family of kets with the property
\begin{equation}
  \int d\avv \, \bigl\|\ket{g(\ptv,\avv)} \bigr\|^2
= 1
\label{eq:  RetOptTCondB}
\end{equation}
for  all $\ptv$.

We conclude this section by showing that for retrodictively
optimal type 1 measurements 
$\langle \am_{\rm i}\rangle= (s+1)\langle  \pt_{\rm f}\rangle$. 
In fact
\begin{align*}
  \bmat{\psi \otimes \ap}{\pt_{\rm f}}{\psi\otimes \ap}
& = \int d\ptv \,\ptv\, \bmat{\psi}{\Sop}{\psi}
\\
& = \frac{2s+1}{4 \pi}
    \int d\ptv \, \ptv \, \bigl| \overlap{\psi}{\ptv}\bigr|^2
\\
& = \frac{1}{s+1} 
    \bmat{\psi\otimes \ap}{\am_{\rm i}}{\psi\otimes \ap}
\end{align*}
where we have used 
the fact~\cite{Lieb} that $(s+1)\ptv$ is the contravariant
symbol corresponding to $\am$.

\section{Predictively Optimal Type 1 Measurements}
\label{sec:  PreOpt}
The purpose of this section is to characterise the form of
the operator $\Top$ and  function $\pdf(\ptv)$ for
processes which maximise the predictive fidelity,
$\pf$.   In the last section we showed that, for retrodictively
optimal type 1 measurements, $\pdf$
coincides with the initial system state $Q$-function.  In
this section we will show that if the measurement is
predictively optimal, then $\pdf$ is
related to the final system state $P$-function.
 
 We begin by
establishing an upper bound on $\pf$.
By a  similar argument to the one leading
to Inequality~(\ref{eq:  RetFidCondC}) we find
\begin{equation*}
  \int d\ptv d\avv \,
     \Tr\bigl( (\pf-\ptv \cdot \am) \,\Top\, \TopD \bigr)
\le 0
\end{equation*}
which only differs from Inequality~(\ref{eq:  RetFidCondC})
in the replacement of $\rf$ by $\pf$, and in the fact that the
order of
$\Top$ and $\TopD$ is reversed.
The analysis therefore proceeds in nearly the same way. 
Corresponding to Inequality~(\ref{eq:  etaiCondA}) we deduce
\begin{equation}
  \pf \le s
\label{eq:  etafCondA}
\end{equation}
which, in view of Eq.~(\ref{eq:  PreErrA}), implies
\begin{equation}
  \ef \amvs  \ge \sqrt{s}
\label{eq:  PreErrRelA}
\end{equation}
We will refer to Inequality~(\ref{eq:  PreErrRelA}) as the
predictive error relation.  It is the analogue, for
measurements of spin direction, of the inequality 
$\PErr x \, \PErr p \ge 1/2$ proved in ref.~\cite{self2b} for
joint measurements of position and momentum (units chosen
such that $\hbar=1$).

We define a predictively optimal type 1 measurement to be one
for which the predictive fidelity is maximal, $\pf =s$ (unlike
the case of retrodictive optimality, we do not impose the
requirement that the measurement also be isotropic).  By a
similar argument to the one given in the last section we 
find, corresponding to  Eqs.~(\ref{eq:  RetOptTCondA})
and~(\ref{eq:  RetOptTCondB}), that the necessary and
sufficient condition for a type 1 measurement to be
predictively optimal is that $\Top$ be of the form
\begin{equation}
  \Top
= \left( \frac{2 s+1}{4 \pi}\right)^{\frac{1}{2}}
  \ket{\ptv} \bra{h(\ptv,\avv)}
\label{eq:  PreOptTCondA}
\end{equation}
for almost all $\ptv$, $\avv$, where $\ket{h(\ptv,\avv)}$ is any
family of kets satisfying the completeness relation
\begin{equation}
   \frac{2 s+1}{4 \pi}
   \int d\ptv d\avv \, \ket{h(\ptv,\avv)} \bra{h(\ptv,\avv)} 
= 1
\label{eq:  PreOptTCondB}
\end{equation}

If $\Top$ is of this form it follows from
Eqs.~(\ref{eq:  SDef})
and~(\ref{eq:  rhoValTermsRhoi}) that
\begin{equation}
 \pdf(\ptv)
= \frac{2 s+1}{4 \pi}
   \int d\avv \,
    \bmat{h(\ptv,\avv)}{\hat{\rho}_{\rm i}}{h(\ptv,\avv)}
\label{eq:  RetroOptPDF}
\end{equation}
where $\hat{\rho}_{\rm i}$ is the initial system  density
matrix.
Now suppose that the measured value of $\ptv$ has been recorded
to lie in the region $\mathscr{R}$ of the unit 2-sphere.  Then,
using Eqs.~(\ref{eq:  rhofTermsT}), (\ref{eq:  PreOptTCondA})
and~(\ref{eq:  RetroOptPDF}), we find
\begin{equation*}
\hat{\rho}_{\rm f}
= \frac{1}{p_{\mathscr{R}}}
  \int_{\mathscr{R}} d\ptv \,
    \pdf (\ptv) \ket{\ptv}\bra{\ptv}
\end{equation*}
where $p_{\mathscr{R}}$ is the probability of recording the
result
$\ptv \in\mathscr{R}$, and where 
$\hat{\rho}_{\rm f}$ is the final system reduced density
matrix.  In view of Eqs.~(\ref{eq:  fCondA}),
(\ref{eq:  fCondB})
and~(\ref{eq:  PfncDef}) this means that the final system state
$P$-function $P_{\rm f}$ is given by
\begin{equation*}
  P_{\rm f} (\ptv)
= \frac{1}{p_{\mathscr{R}}}
  \int_{\mathscr{R}} d\ptv' \,
    \Pi_{2s}(\ptv,\ptv') \pdf(\ptv')
\end{equation*}
for almost all $\ptv$, where $\Pi_{2s}$ is the projection kernel
defined by Eq.~(\ref{eq:  Pi0Proj}).

If $\mathscr{R}$ is a sufficiently small region surrounding  the
point
$\ptv_{0}$ then
\begin{equation*}
  \hat{\rho}_{\rm f} \approx \ket{\ptv_{0}}\bra{\ptv_{0}}
\end{equation*}

Finally, we note that for a predictively optimal type 1
measurement 
$\langle \am_{\rm f} \rangle= s\langle \pt_{\rm f}\rangle$.  In
fact
\begin{align*}
  \bmat{\psi \otimes \ap}{\am_{\rm f}}{\psi \otimes \ap}
& = \int d\ptv d\avv \,
     \bmat{\psi}{\TopD \,  \am \, \Top}{\psi}
\\
& =  \frac{2 s+1}{4 \pi}\int d\ptv d\avv\bmat{\ptv}{\am}{\ptv}
         \boverlap{\psi}{h(\ptv,\avv)}
         \boverlap{h(\ptv,\avv)}{\psi}
\\
& =  s \int d\ptv \,  \ptv 
       \bmat{\psi}{\Sop}{\psi}
\\
& =  s  \bmat{\psi \otimes \ap}{\pt_{\rm f}}{\psi \otimes \ap}
\end{align*}
where we have used the fact~\cite{Lieb} that $s \ptv$ is the
covariant symbol corresponding to $\am$.

\section{Completely Optimal Type 1 Measurements}
\label{sec:  CompOpt}
We define a completely optimal type 1 measurement to be one
which is both retrodictively and predictively optimal.
Referring to  Eqs.~(\ref{eq:  RetOptTCondA}),
(\ref{eq:  RetOptTCondB}),
(\ref{eq:  PreOptTCondA}),
and~(\ref{eq:  PreOptTCondB}) we see that the necessary and
sufficient condition for this to be true is that
$\Top$ be of the form
\begin{equation*}
  \Top
= \left(\frac{2 s+1}{4\pi}\right)^{\frac{1}{2}} f(\ptv, \avv)
   \ket{\ptv}\bra{\ptv}
\end{equation*}
where $f$ is any function with the property
\begin{equation*}
  \int d \avv \, \left| f(\ptv, \avv)\right|^2 =1 
\end{equation*}
for  all $\ptv$.

Expressed in terms of the operator $\hat{U}$ the condition
reads [see Eq.~(\ref{eq:  TDef})]
\begin{equation*}
  \bigl( \bra{m_1}\otimes\bra{\ptv,\avv} \bigr)
  \hat{U}
  \bigl(  \ket{m_2} \otimes \apst\bigr)
=  \left(\frac{2 s+1}{4\pi}\right)^{\frac{1}{2}} f(\ptv, \avv)
   \overlap{m_1}{\ptv} \overlap{\ptv}{m_2}
\end{equation*}
It is straightforward to  verify that there do exist
unitary operators
$\hat{U}$ with this property.  It follows that completely
optimal measurements are defined mathematically.  The question
as to whether they are possible physically is, of course, rather
less straightforward.

Referring to Eq.~(\ref{eq:  etadTermsT}) we see that, for a
completely optimal measurement, the quantity $\syf$,
characterising the extent to which the system is disturbed by
the measurement process, is given by
\begin{equation}
\syf
 = \inf_{\ket{\psi}\in\mathscr{S}_{\rm sy}}
  \biggl( \frac{2 s+1}{4 \pi}
   \int d \ptv  \, \frac{s}{2} \Bigl( 
     \boverlap{\psi}{\ptv} \bmat{\ptv}{\ptv \cdot \am}{\psi}
     + \bmat{\psi}{\ptv \cdot \am}{\ptv}\boverlap{\ptv}{\psi}
     \Bigr)
     \biggr)
= s^2
\label{eq:  ComOptetad}
\end{equation}
where we have used the fact~\cite{Lieb} that $s\ptv$ is the
covariant symbol corresponding to 
$\am $.  In view of Eq.~(\ref{eq:  DistA}) it follows that
\begin{equation*}
  \dis \amvs = \sqrt{2 s}
\end{equation*}

\section{Type 2 Measurements}
\label{sec:  type2}
In the preceding sections we have been concerned with type 1
measurements, for which the pointer position is constrained to
lie on the unit $2-$sphere.  We now turn our attention to type
2 measurements.  As explained in the Introduction, these are
measurements for which the outcome is represented by the three
\emph{independent} commuting components of a vector $\mpt$, no
constraint being placed on the squared modulus
$\mpts^2=\sum_{a=1}^{3} \mpts^{2}_{a}$.  We will show that, the
more nearly a type 2 measurement approaches to optimality, the
more nearly it approximates  an (optimal) type 1
measurement. 

We first need to characterise the accuracy of a type 2
measurement.   A  similar analysis to that given in 
Section~\ref{sec:  POVM} can be carried through for  type 2
measurements, with the replacement
$\ptv\rightarrow\mptv$.
As before, we denote the additional apparatus
degrees of freedom $\av=(\av_{1},\dots,\av_{N})$, so that the
eigenkets $\ket{\mptv,\avv}$ comprise an orthonormal basis for
the apparatus state space, $\mathscr{H}_{\rm ap}$.  Let $\apst$
be the intial apparatus state, and let $\hat{U}$ be the unitary
operator describing the evolution brought about by the
measurement interaction.  Then, if the initial system
state is
$\ket{\psi}$, the final state of system$+$apparatus,
immediately after the measurement interaction has ended, will
be  given by
$\hat{U}\ket{\psi \otimes \ap}$.  Corresponding
to Eqs.~(\ref{eq:  TDef})  and~(\ref{eq:  SDef}) we define
\begin{equation*}
\Topm
=
\sum_{m,m'=-s}^{s}
 \bigl(\bra{m}\otimes
             \bra{\mptv,\avv}\bigr)
          \hat{U}
          \bigl(\ket{m'}\otimes \apst \bigr)
          \ket{m}\bra{m'}
\end{equation*}
and
\begin{equation}
  \Sopm
=  \int d \avv \, \TopmD \, \Topm
\label{eq:  EType2Def}
\end{equation}
Corresponding to Eqs.~(\ref{eq:  RetErrA}) 
and~(\ref{eq:  PreErrA}) we define the
maximal rms errors of retrodiction and prediction by
\begin{align}
\ei \amvs 
&  = \left(\sup_{\ket{\psi }\in \mathrm{S}_{\rm sy}}
  \left( \bmat{\psi\otimes\ap}{
               \bigl| \mpt_{\rm f}-\am_{\rm i}
\bigr|^2}{\psi\otimes\ap}
  \right)
        \right)^{\frac{1}{2}}
\label{eq:  RetErrB}
\\
\intertext{and}
\ef \amvs 
&  = \left(\sup_{\ket{\psi }\in \mathrm{S}_{\rm sy}}
  \left( \bmat{\psi\otimes\ap}{
               \bigl|\mpt_{\rm f}-\am_{\rm f}
\bigr|^2}{\psi\otimes\ap}
  \right)
        \right)^{\frac{1}{2}}
\label{eq:  PreErrB}
\end{align}
where $\am_{\rm i}=\am$, 
$\am_{\rm f}=\hat{U}^{\dagger} \am \hat{U}$
and $\mpt_{\rm f} = \hat{U}^{\dagger} \am \hat{U}$.
It can be seen that Eq.~(\ref{eq:  RetErrB})
agrees with Eq.~(\ref{eq:  RetErrA}) if one replaces
$\mpt_{\rm f} \rightarrow \rf \pt_{\rm f}$, and that 
Eq.~(\ref{eq:  PreErrB}) agrees with 
Eq.~(\ref{eq:  PreErrA}) if one replaces
$ \mpt_{\rm f} \rightarrow \pf \pt_{\rm f}$.

In terms of the operators $\Sopm$  and $\Topm$ we have
\begin{equation}
  \ei \amvs
 = 
  \Biggl(\sup_{\ket{\psi}\in\mathscr{S}_{\rm sy}}
  \biggl( \int d \mptv \,
     \sum_{a=1}^{3}
     \bmat{\psi}{
     (\mptvs_{a}-\ams_{a}) \, \Sopm \, (\mptvs_{a}-\ams_{a})}{
     \psi} \biggr) \Biggr)^{\frac{1}{2}}
\label{eq:  RetErrType2Def}
\end{equation}
and
\begin{equation}
  \ef \amvs
 = 
  \Biggl(
  \sup_{\ket{\psi}\in\mathscr{S}_{\rm sys}}
  \biggl( \int d \mptv  d \avv \,
    \bmat{\psi}{\hat{T}^{\dagger}(\mptv,\avv) \,
      \bigl|\mpt - \am\bigr|^2 \,
      \hat{T}(\mptv,\avv)}{\psi}
  \biggr)\Biggr)^{\frac{1}{2}}
\label{eq:  PreErrType2Def}
\end{equation}

We next show that, corresponding to 
Inequality~(\ref{eq:  RetErrRelA}), one has the retrodictive 
error relationship for type
2  measurements
\begin{equation}
  \ei \amvs \ge \sqrt{s}
\label{eq:  RetErrRelB}
\end{equation}
and that, corresponding to
Inequality~(\ref{eq:  PreErrRelA}), one has the 
predictive error relationship for type
2  measurements
\begin{equation}
  \ef \amvs  \ge \sqrt{s}
\label{eq:  PreErrRelB}
\end{equation}
In fact, it follows from Eqs.~(\ref{eq:  EType2Def}),
(\ref{eq:  RetErrType2Def})
and~(\ref{eq:  PreErrType2Def}) that
{\allowdisplaybreaks
\begin{align*}
(2 s+1) \left(\ei \amvs\right)^2 
& \ge
 \int d\mptv d\avv \,
 \Tr \left(\bigl|\mptv-\am\bigr|^2 \, \TopmD \, \Topm \right)
\\
\intertext{and}
(2 s+1) \left(\ef \amvs \right)^2 
& \ge
 \int d\mptv  d\avv \,
 \Tr \left(\bigl|\mptv-\am\bigr|^2  \, \Topm\, \TopmD\right)
\end{align*}
Using the fact
\begin{equation*}
  (2 s+1) = \int d\mptv d\avv \, 
    \Tr\left(\TopmD \, \Topm\right)
\end{equation*}
we deduce
\begin{align}
\int d\mptv d\avv \,
 \Tr \biggl(\left(\bigl|\mptv-\am\bigr|^2 -
           \left(\ei \amvs\right)^2 
           \right)
          \TopmD\,    \Topm \biggr)
& \le 0
\label{eq:  RetErrBCondA}
\\
\intertext{and}
\int d\mptv d\avv \,
 \Tr \biggl(\left(\bigl|\mptv-\am \bigr|^2 -
           \left(\ef \amvs\right)^2 
           \right)
          \Topm \, \TopmD\biggr)
& \le 0
\label{eq:  PreErrBCondA}
\end{align}
Now make the expansion}
\begin{equation*}
  \Topm
= \sum_{m,m'=-s}^{s}
   T_{m m'} (\mptv,\avv) \ket{\ptv,m}\bra{\ptv,m'}
\end{equation*}
where $\ptv = \mptv / \mptvs$ and  
$\ket{\ptv,m}$ is the state defined by
Eq.~(\ref{eq:  mnKetDef}).
Using this expansion Inequalities~(\ref{eq:  RetErrBCondA})
and~(\ref{eq:  PreErrBCondA}) become
\begin{align}
 \sum_{m,m'=-s}^{s} 
 \int d\mptv d\avv \,
 \Bigl(\bigl(\mptvs-m'\bigr)^2+\bigl(s^2
  - m'\vphantom{s}^2\bigr) 
    +\bigl(s-(\ei \amvs)^2 \bigr) 
           \Bigr)
 \bigl| T_{m m'}(\mptv,\avv) \bigr|^2
& \le 0
\label{eq:  RetErrBCondB}
\\
\intertext{and}
 \sum_{m,m'=-s}^{s} 
 \int d\mptv d\avv \,
 \Bigl(\bigl(\mptvs-m\bigr)^2+\bigl(s^2
  - m^2\bigr) 
    +\bigl(s-(\ef \amvs)^2 \bigr) 
           \Bigr)
 \bigl| T_{m m'}(\mptv,\avv) \bigr|^2
& \le 0
\label{eq:  PreErrBCondB}
\end{align}
Inequalities~(\ref{eq:  RetErrRelB})
and~(\ref{eq:  PreErrRelB}) are now immediate.

Setting $\ei \amvs = \sqrt{s}$ in 
Inequality~(\ref{eq:  RetErrBCondB}) gives
\begin{equation*}
 \sum_{m,m'=-s}^{s} 
 \int d\mptv d\avv \,
 \Bigl(\bigl(\mptvs-m'\bigr)^2+\bigl(s^2
  - m'\vphantom{s}^2\bigr)  
           \Bigr)
 \bigl| T_{m m'}(\mptv,\avv) \bigr|^2
\le 0
\end{equation*}
which implies 
\begin{equation*}
  \left| T_{m m'}(\mptv,\avv)\right|^2
= g_{m}(\ptv,\avv) \, \delta_{m' s} \, \delta(\mptvs -s)
\end{equation*}
for suitable functions $g_{m}$.  However, this is not possible,
since the square root of the $\delta$-function is not defined.
It follows that the lower bound set by 
Inequality~(\ref{eq:  RetErrRelB}) is not precisely
achievable.  Nor is the lower bound set by
Inequality~(\ref{eq:  PreErrRelB}).

It is, however, possible to approach the lower bounds set by
Inequalities~(\ref{eq:  RetErrRelB})
and~(\ref{eq:  PreErrRelB}) arbitrarily closely.  It can be
seen that as $\ei \amvs \rightarrow \sqrt{s}$ (respectively,
$\ef
\amvs
\rightarrow \sqrt{s}$), then 
$\Topm$ and $\Sopm$ become more and more
strongly concentrated on the surface $\mptvs =s$.  In other
words, the measurement more and more nearly approaches
a type 1 measurement of maximal retrodictive (respectively,
predictive) accuracy, with pointer observable
$\pt=\mpt/s$.
\section{Conclusion}
\label{sec:  conclusion}
There are a number of ways in which one might seek to develop
the results reported in this paper.

In the first place, although we showed that 
$\dis \amvs = \sqrt{2 s}$ for a completely optimal type 1
measurement, we did not derive error-disturbance relationships,
analogous to the inequalities $\ei x \,\dis p$, $\ei p\, \dis x$,
$\ef x \, \dis p$, $\ef p \, \dis x \ge 1/2$ (in units such
that $\hbar=1$) proved in ref.~\cite{self2b} for the case of a
simultaneous measurement of position and momentum.  The general
principles of quantum mechanics~\cite{Heisenberg,Braginsky}
indicate that  relationships of this kind must also hold for
measurements of spin direction, at least on a qualitative
level.  However, it appears that the problem of giving the
relationships precise, numerical expression is not entirely
straightforward.  The question requires further investigation.

In this paper we have considered measurements of spin
direction.  However, the problem of simultaneously measuring
just two components of spin is also
important~\cite{TwoComp,BuschSpin}.  It would be interesting to
investigate the accuracy of measurements such as this, and to
try to characterise the POVM (or POVM's, in the plural?)
describing the outcome when the measurement is optimal.

We have seen that $\mathrm{SU}(2)$ coherent states play an
important role in the description of optimal measurements of
spin direction.   In refs.~\cite{self3,Ali} it was
shown that ordinary, Heisenberg-Weyl coherent states play an
analogous role in the description of optimal joint measurements
of position and momentum.  It would be interesting to see if it
is generally true, that every system of generalized coherent
states is related in this way to  joint measurements of
the generators of the corresponding Lie group.

There are some important questions of principle regarding
measurements of a \emph{single} spin
component~\cite{Busch,BuschSpin,Wigner,Wheeler,Garra}.  It would
be interesting to see if the approach to the problem of defining
the measurement accuracy which was described in this paper can
be used to gain some additional insight into these questions.

Finally, it is obviously important to investigate
whether optimal, or near optimal determinations of spin
direction can be realised experimentally.

\end{document}